\documentclass[conference]{IEEEtran}
\IEEEoverridecommandlockouts
\usepackage{cite}
\usepackage{hyperref}
\hypersetup{
    colorlinks=true,
    linkcolor=black,
    citecolor=black,
    filecolor=black,
    urlcolor=black,
}
\usepackage{amsmath,amssymb,amsfonts}
\usepackage{algorithmic}
\usepackage{graphicx}
\usepackage{textcomp}
\usepackage{xcolor}
\usepackage{tabularx}
\usepackage{multirow}
\usepackage{dirtytalk}
\usepackage{pgf-pie}
\usepackage{caption}
\usepackage{booktabs}
\usepackage{float, balance}
\usepackage{url}
\usepackage{pdfpages}

\definecolor{superlight}{rgb}{0.949, 0.949, 0.949}
\renewcommand\fbox{\fcolorbox{black}{superlight}}

\usepackage[most]{tcolorbox}
\tcbset{textmarker/.style={%
        enhanced, breakable,
        parbox=false,boxrule=0mm,boxsep=0mm,arc=0mm,
        outer arc=0mm,left=1.8mm,right=1.8mm,top=7pt,bottom=7pt,
        toptitle=1mm,bottomtitle=1mm,oversize}}

\newtcolorbox{noteBox}{
    breakable,
    enhanced,
    textmarker,
    borderline west={3pt}{0pt}{gray},
    colback=gray!10!white
}

\usepackage{framed}
\definecolor{formalshadelight}{RGB}{242,242,242}
\definecolor{formalshadedark}{RGB}{166,166,166}

\def\BibTeX{{\rm B\kern-.05em{\sc i\kern-.025em b}\kern-.08em
    T\kern-.1667em\lower.7ex\hbox{E}\kern-.125emX}}
\begin{document}

\title{Exploring Technical Debt in Security Questions on Stack Overflow
% \thanks{978-1-6654-5223-6/23/\$31.00 ©2023 IEEE}
}

% \author{\IEEEauthorblockN{Joshua Aldrich Edbert}
% \IEEEauthorblockA{\textit{Department of Computer Science} \\
% \textit{University of Saskatchewan}\\
% % Saskatoon, Canada \\
% joshua.edbert@usask.ca}
% \and
% \IEEEauthorblockN{Sahrima Jannat Oishwee}
% \IEEEauthorblockA{\textit{Department of Computer Science} \\
% \textit{University of Saskatchewan}\\
% % Saskatoon, Canada \\
% sao107@usask.ca}
% \and
% \IEEEauthorblockN{Shubhashis Karmakar}
% \IEEEauthorblockA{\textit{Department of Computer Science} \\
% \textit{University of Saskatchewan}\\
% % Saskatoon, Canada \\
% shk106@usask.ca}
% \and
% \IEEEauthorblockN{Zadia Codabux}
% \IEEEauthorblockA{\textit{Department of Computer Science} \\
% \textit{University of Saskatchewan}\\
% % Saskatoon, Canada \\
% zadiacodabux@ieee.org}
% \and
% \IEEEauthorblockN{Roberto Verdecchia}
% \IEEEauthorblockA{\textit{Department of Information Engineering} \\
% \textit{University of Florence}\\
% % Florence, Italy \\
% roberto.verdecchia@unifi.it}

% }

\author{
\IEEEauthorblockN{
Joshua Aldrich Edbert\IEEEauthorrefmark{1}, 
Sahrima Jannat Oishwee\IEEEauthorrefmark{1}, 
Shubhashis Karmakar\IEEEauthorrefmark{1}, 
Zadia Codabux\IEEEauthorrefmark{1}, 
Roberto Verdecchia\IEEEauthorrefmark{2}}
\IEEEauthorblockA{
\IEEEauthorrefmark{1}Department of Computer Science\\
University of Saskatchewan\\
Email: \{joshua.edbert, sahrima.oishwee, shubhashis.k\}@usask.ca, zadiacodabux@ieee.org}
\IEEEauthorblockA{
\IEEEauthorrefmark{2}Department of Information Engineering\\
University of Florence\\
Email: roberto.verdecchia@unifi.it} 
% \IEEEauthorblockA{
% \IEEEauthorrefmark{3}
% Starfleet Academy, San Francisco, California 96678-2391\\
% Telephone: (800) 555--1212, Fax: (888) 555--1212}
% \IEEEauthorblockA{
% \IEEEauthorrefmark{4}
% Tyrell Inc.,123 Replicant Street, Los Angeles, California 90210
% --4321}
}

\maketitle
\IEEEpeerreviewmaketitle 

\begin{abstract}

Background: Software security is crucial to ensure that the users are protected from undesirable consequences such as malware attacks which can result in loss of data and, subsequently, financial loss. Technical Debt (TD) is a metaphor incurred by suboptimal decisions resulting in long-term consequences such as increased defects and vulnerabilities if not managed. Although previous studies have studied the relationship between security and TD, examining their intersection in developers' discussion on Stack Overflow (SO) is still unexplored. 
Aims: This study investigates the characteristics of security-related TD questions on  SO. More specifically, we explore the prevalence of TD in security-related queries, identify the security tags most prone to TD, and investigate which user groups are more aware of TD.
Method: We mined 117,233 security-related questions on SO and used a deep-learning approach to identify 45,078 security-related TD questions. Subsequently, we conducted quantitative and qualitative analyses of the collected security-related TD questions, including sentiment analysis.
Results: Our analysis revealed that 38\% of the security questions on SO are security-related TD questions. The most recurrent tags among the security-related TD questions emerged as ``security" and ``encryption." The latter typically have a neutral sentiment, are lengthier, and are posed by users with higher reputation scores.
Conclusions: 
Our findings reveal that developers implicitly discuss TD, suggesting developers have a potential knowledge gap regarding the TD metaphor in the security domain. Moreover, we identified the most common security topics mentioned in TD-related posts, providing valuable insights for developers and researchers to assist developers in prioritizing security concerns in order to minimize TD and enhance software security.

%optimal prevention techniques. Therefore, further investigation into developers' difficulties in technical debt resolution in security questions needs to be conducted to minimize incurring technical debt without compromising software security. 

% By comprehending the link between technical debt and security, developers may focus resources and efforts on preventing and reducing technical debt in the security topics more prone to technical debt, creating safer and easier-to-manage software systems. Additionally, these results help direct researchers and practitioners to focus on security areas most vulnerable to technical debt and raise developer awareness of the presence of technical debt in security.
\end{abstract}


\section{Introduction}

% \textcolor{blue}{am adding a comment here so that you all think about it early on. a few times in the paper, it's written "TD discussion." and I've left few comments but i think it's a big issue if left uncorrected in the paper. I don't think you're exploring TD in security posts. it's like looking for specific instances of code debt, design debt, code smells in security posts. But rather you are looking for anything that is "sub optimal" in the security questions. so there is a difference. so far, in the paper, you write as if you are doing the first one when you are in fact doing the second one. I don't know how you want to phrase it. it could be ``suboptimal security solutions in the security questions" - you wrote "``suboptimal security solutions" in the second column of the intro - but is it always solutions that are discussed in the questions? You have to judge what's the best terminology to use since you looked at the questions. one thing that you can do is explain upfront in intro that what you mean by TD in the context of this study - "all the unoptimal solutions, ...... in the security questions." make sure you decsribe it properly, then you can use TD throughout the paper, else as is, it is flawed.}

Security is a crucial component in software development~\cite {wen2019learning} and is concerned with the ability of applications to withstand malicious attacks brought on by exploiting flaws in the software~\cite{assal2018security}. Neglecting security in software development will result in severe consequences for users and companies, such as financial loss, personal data compromise, confidentiality breaches, damaged reputation, and delays in software development efforts~\cite{yang2016security, moradian2013security, khan2022systematic, khan2011secure}. In addition, security threats in software development have become more complex due to growing advancements in the Internet of Things~\cite{michael2006security}. Hence, ensuring secure software is crucial to software development~\cite{geer2010companies, tondel2008security}.

Technical Debt (TD) is a metaphor used to describe suboptimal artifacts created as a result of design and implementation choices that, while they may achieve short-term objectives, may cause issues during the maintenance and evolution phases of a software project~\cite{kruchten2012technical}. TD can be incurred during any phase of the software development lifecycle but primarily impacts software maintenance~\cite{alves2016identification}. Inadequate TD management can result in higher costs, poor product quality, and a slowdown in the long-term success of software development~\cite{lim2012balancing}. As a result, TD is also acknowledged as a critical problem in software development~\cite{tom2013exploration}.

Identifying security threats during the software development lifecycle is crucial for secure development\cite{mcgraw2012software}. Hence, mechanisms to help developers detect security risks before software releases are important~\cite{siavvas2018static, siavvas2019optimum}. 
For instance, TD has been used to highlight security vulnerabilities in software products~\cite{siavvas2019empirical}. Suboptimal security implementation practices can weaken a system disastrously, and simple coding mistakes or design issues can lead to exploitable vulnerabilities~\cite{kuhn2018can, camilo2015bugs}. Incurring TD due to suboptimal security practices can make it harder to maintain or update the software, increasing the risk of future attacks and further compounding the consequences of a breach~\cite{siavvas2019empirical}. Therefore, studying the intersection between TD and security in developers' discussions is essential to understanding and preventing suboptimal security practices.

Previous studies have explored the connection between TD and security vulnerability~\cite{siavvas2019empirical, izurieta2018position, siavvas2022technical, rindell2019security} and examined the security-related posts on Stack Overflow (SO)~\cite{yang2016security}. Additionally, studies have examined the TD-related posts on SO~\cite{alfayez2023asked, asking_about_technical_debt}. However, to the best of our knowledge, no studies have focused explicitly on security-related questions with TD on SO. SO is a crowdsourcing platform where users exchange information about programming tasks with over 24 million questions and 20 million users\footnote{\url{https://stackexchange.com/sites\#traffic}}. It allows practitioners to examine how users request support and exchange expertise about technical issues~\cite{asking_about_technical_debt}. SO is a helpful resource for understanding real-world viewpoints on various software engineering problems~\cite{barua2014developers, silva2021topic}. 
% Thus, insights into the difficulties experienced by developers in the context of TD and security can be gained from analyzing questions on this platform, pointing out areas that require additional study and development. 

% Analyzing Stack Overflow questions can provide insights into the relationship between security and TD and get structured knowledge that cannot be learned solely through personal experience. This knowledge can assist developers in creating more secure software.

In this study, we investigated security-related questions on SO. The security-related questions with indications of suboptimal security practices will be referred to as Security-related TD Questions (STDQs). The security questions without indications of suboptimal security practices will be referred to as non-STDQs. More specifically, we quantitatively and qualitatively investigated various traits of STDQs and non-STDQs, including the security tags that are more recurrent in STDQs and non-STDQs, the sentiment, popularity score, length, security topics, question types of STDQs and non-STDQs, and finally, the SO account profile of users asking STDQs and non-STDQs. Note that security ``tag" and ``topic" are different terms used in this study. Tags are SO tags attached to the questions by the users, while topics emerged from our manual coding process of the questions (described in Section \ref{subsec:dataanalysis}). 

% This study can provide insights into the relationship between TD and security for developers in creating more secure software.}
 
This study offers a thorough analysis of STDQs to help better comprehend the relationship between TD and software security. Our key contributions include the following:
\begin{enumerate}
    \item A qualitative analysis of question types and security topics of STDQs to better inform the community which specific question types and security topics are more prevalent in STDQs.
	% \item \textcolor{red}{An analysis of TD in security-related questions highlights the significance of TD in developing secure software.}
    \item An exploration of the most recurrent security tags in STDQs to support researchers in identifying the areas of security most susceptible to TD.
    \item An analysis of the question length, sentiment, popularity, and time needed for answers of STDQs to understand the complexity, popularity, and emotion toward STDQs.
    \item An investigation of the SO account profile characteristics of the users asking STDQs to understand which group of people most frequently pose STDQs and need more understanding of TD resolution in secure development.
	\item A comprehensive replication package\footnote{\url{https://doi.org/10.5281/zenodo.7888440}} of the study.
\end{enumerate}

The rest of this paper is structured as follows: Section \ref{sec:relevant_works} presents related works. Section \ref{sec:methodology} describes our study's methodology. Section \ref{sec:findings} and Section \ref{sec:discussion} describe and discuss the findings of our experiments, respectively. Section \ref{sec:implications} describes the implications of our findings. Threats to this study's validity are listed in Section \ref{sec:threats_to_validity}. Section \ref{conclusion} concludes the study. \\

% \section{Background}

% Stack Overflow (SO) is an online platform where developers can ask programming-related questions, discuss related topics, and find solutions to their problems \cite{yang2016security}. To help other users understand the question, a developer must add tags before publishing a question. A user's response is considered the accepted answer to a question if it provides a solution to the questioner's issue and is chosen by the questioner. 

% Users can vote on both the questions and the answers on the website. The terms "upvote" and "downvote" indicate how helpful a question or response was to other users. The positive votes are referred to as upvotes, and the negative votes are known as downvotes.  

% A user's reputation score on Stack Overflow is based on various actions, including question-and-answer posting, answer voting, and commenting. These actions help users establish a positive reputation on the SO website. The greater a member's reputation score, the more options they have, including the ability to close inquiries and remove questions and replies.

% Since its launch in 2008, SO has gradually gained recognition among developers. When this paper was being written, SO had 34 million answer-type answers for the questions and 23 million question-type posts. In SO, 70\% of questions have responses. Eighteen million people are active users of SO as of November 2022.

\section{Related Work}
\label{sec:relevant_works}
% While there are studies investigating various aspects of software development, we focus on SO studies on TD (Section \ref{RW:TD-SO}) and security (Section \ref{RW:Security-SO}). \textcolor{red}{Section \ref{RW:general-SO} summarizes general SO studies, and Section \ref{RW:security-TD} describes general studies on security and TD.}

\textbf{General Stack Overflow Studies.}
\label{RW:general-SO}
Barua et al.~\cite{barua2014developers} examined the main topics and trends of general SO discussions using Latent Dirichlet Allocation. Although our study and theirs analyzed SO discussions, our study did not investigate general SO discussions but investigated STDQs specifically. Their study found that the topic of interest among developers varies. Over time, web development, mobile applications, Git, and MySQL have gained the most traction. Rosen et al.~\cite{rosen2016mobile} analyzed the topics related to mobile development in SO discussions instead of TD and security. Their study revealed that questions like app distribution, mobile APIs, and data management are frequently asked in SO. Haque et al.~\cite{haque2020challenges} also examined the topics of SO discussions related to Docker instead of security and TD. According to their findings, most developers use SO to post questions about various Docker-related subjects, such as framework development and application deployment.

% \subsection{\textcolor{red}{Stack Overflow Studies on Technical Debt}}
\textbf{Stack Overflow Studies on Technical Debt.}
\label{RW:TD-SO}
Kozanidis et al.~\cite{asking_about_technical_debt} analyzed the different characteristics of TD-related questions on SO. This study is very similar since we also analyzed the characteristics of SO questions, including sentiment and question length. However, we analyzed STDQs instead of TD-related questions. In addition, they investigated whether machine learning can be used to detect and classify TD questions on SO automatically. We used their dataset for training our classification model to detect STDQs. Their study showed that architecture debt is the most discussed debt on SO, most TD questions have a slight sense of urgency and neutral sentiment, the question length varies across different debt types, and machine learning can identify TD questions but not classify them across different debt kinds. The study by Alfayez et al.~\cite{alfayez2023asked} extracted and examined 578 TD-related queries using a dataset derived from three Stack Exchange Q\&A websites, including SO. Although their study and ours investigated TD-related questions on SO, they did not specifically investigate STDQs. In addition, most of the characteristics we investigated differed from theirs, except for the median time to receive an answer. Their findings showed that there are 14 categories in which TD-related questions can be categorized, 636 different tags are used in the acquired set of TD-related questions, and some TD-related categories have a shortage of accepted answers and a longer median time to receive an accepted answer than others. Gama et al.~\cite{gama2020using} manually evaluated a sample of SO discussions to determine how developers identify TD in their software projects. While their study and ours investigated SO, we aimed to investigate STDQs, not TD identification, in software projects. They found that SO users frequently discuss TD identification and reported 29 low-level indicators for detecting TD items in code, infrastructure, architecture, and test.

% Digkas et al.~\cite{digkas2019reusing} looked into the relationship between using previously published code from Stack Overflow and having TD. They used source code static analysis to examine the TD caused by reusing code snippets offered on the website. Reused code snippets generally exhibit a significantly lower TD density than the host project. In most cases, the reused code has no TD items. 

% Another study about TD and Stack Overflow concerns Perez et al.~\cite{perez2019proposed} model-driven approach to managing the architectural TD life cycle. This study evaluates the sentiment expressed while documenting architectural design decisions using a dataset of Stack Overflow posts~\cite{calefato2018sentiment}.With an emphasis on the architectural level, this approach aims to assist in making clear the impact of the incurring architectural TD.

\textbf{Stack Overflow Studies on Software Security.}
\label{RW:Security-SO}
Yang et al.~\cite{yang2016security} described a large-scale study on security-related questions on SO. Their study investigated security-related topics and trends. Their study and ours investigated security-related questions on SO, but ours specifically investigated STDQs. The authors used the SO tagging system to extract security-related questions from SO, which we used in our study to collect the security-related questions of SO. They reported the top five main security-related topics and the top eight challenging security-related topics. 
Croft et al.~\cite{croft2022empirical} examined SO and GitHub discussions. 
% They thoroughly investigated the security issues faced by programmers of 15 languages. They leveraged the Latent Dirichlet Analysis as a methodology for the study. According to their research, the difficulties and their features vary greatly depending on the programming languages and data sources. 
They investigated the security issues faced by programmers using different programming languages by leveraging  Latent Dirichlet Analysis. The difficulties and features of the security issues vary greatly depending on the programming languages and data sources. 
Lopez et al.~\cite{lopez2019anatomy} studied how security knowledge and secure practices were produced and shared among practitioners on SO and how developers conversed about security in SO. Their results showed developers actively conversed on the website to address security issues, promote knowledge, exchange knowledge, and help one another.

% Ferreyra et al.~\cite{ferreyra2022cybersecurity} analyzed users' engagement and privacy practices in SO. The study focused on cybersecurity-related discussions, while our study analyzed security questions generally. They explicitly examined developers' self-disclosure behavior regarding profile visibility and participation in privacy and security-related conversations. They reported that proactive users share much less information in their account profile than reactive and disengaged users. Nevertheless, there are no associations between these engagements and privacy-related concepts.

% Our study did not examine GitHub discussions and did not focus solely on security issues related to programming languages. 

% They divided all security-related topics into five major areas, with web security accounting for most questions. The most famous four security-related subjects are ``Password," ``Hash," ``Signature," and ``SQL Injection." Java Security, Asymmetric Encryption, Bug, Browser Security, Windows Authority, Signature, ASP.NET, and Password are the top eight challenging security-related topics. 

\textbf{Security and Technical Debt Studies}
\label{RW:security-TD}
Siavvas et al.~\cite{siavvas2019empirical} assessed TD's ability to detect security issues in software products by analyzing a large code repository with static analysis tools. They found a statistically significant positive correlation between TD and the vulnerability densities of the examined software products. Siavvas et al.~\cite{siavvas2022technical} evaluated TD indicators in predicting software security risks at the project and class levels by developing various machine learning models. Their conclusions imply that TD indications have the potential to serve as security indicators. Izurieta et al.~\cite{izurieta2018position} developed a method to examine TD-related security flaws using the Common Weakness Enumeration and the Common Weakness Scoring System. 
While our study also investigated the intersection between TD and security, we specifically focused on SO discussion, an aspect not explored by previous studies.

\textbf{Summary}: Compared to previous studies, which investigated either TD or security vulnerability discussions in SO separately, our study specifically investigated STDQs. We explored the user characteristics such as the SO account age, reputation score, profile views, and earned badges. Our study also identified security topics more prone to TD. \\

\section{Methodology}
\label{sec:methodology}
\begin{figure*}[!ht]
\centering
\includegraphics[width=12cm, height=7cm]{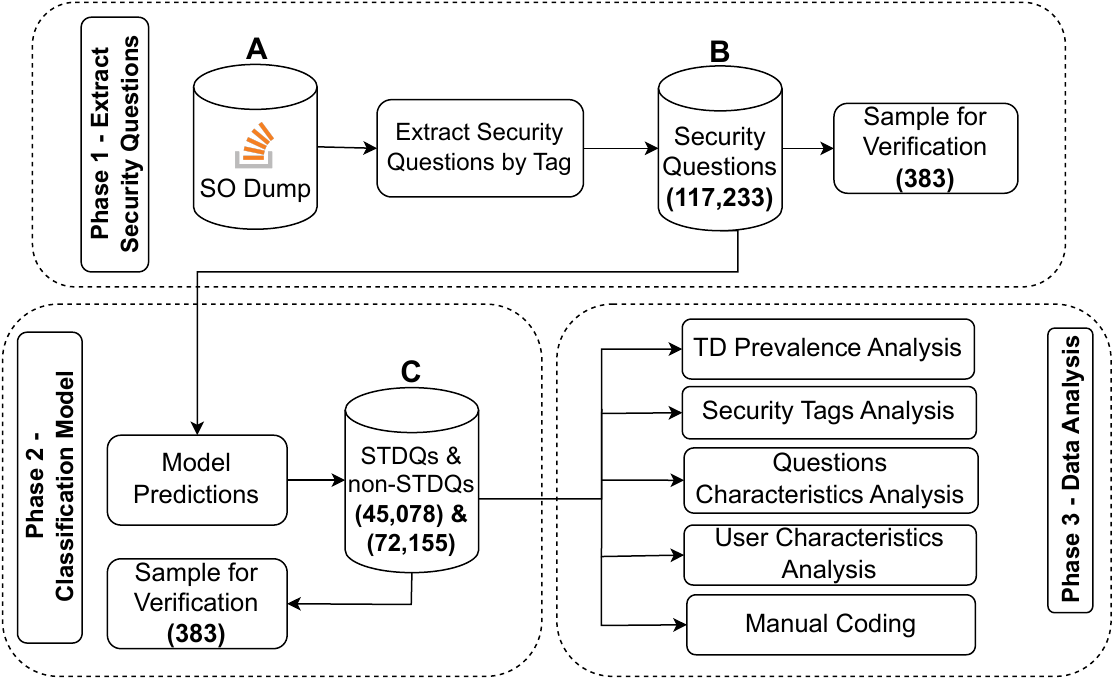}
\caption[Caption for fig1]{Methodology Overview}
\label{fig:methodology}
\end{figure*}

This section describes how this study was conducted. We describe the objective, research questions, data collection, and analysis. \autoref{fig:methodology} depicts the phases of the methodology.

% \vspace{-2px}

\subsection{Goal}
This goal of the study is described using the Goal-Question-Metric technique~\cite{gqm} as follows:
\begin{quote}
\textbf{Purpose:} To investigate\\
\textbf{Issue:} the characteristics of\\
\textbf{Object:} security-related technical debt questions\\
\textbf{Viewpoint:} from the software engineering researchers perspective
\end{quote}

\subsection{Research Questions}
Based on our goal, we derive the following Research
Questions (RQs):

\textbf{$\mathbf{RQ_1}$ To what extent do developers indicate suboptimal security practices in security-related questions?}

\textit{Rationale: }This research question seeks to understand how frequently suboptimal security practices are indicated in security-related questions on SO. Understanding this will help us gain insights into the prevalence of incurring suboptimal security practices in security questions. This research question contributes to the broader understanding of how TD and security are intertwined from the SO discussion perspective. 

\textbf{$\mathbf{RQ_2}$ Which security tags are more recurrent in security-related TD questions?}

\textit{Rationale: }By exploring which security-related tags on SO contain the most indications of suboptimal security practices, researchers can identify the areas of security most susceptible to TD. Identifying these areas can help software developers prioritize their resources and efforts to prevent and mitigate TD in the identified areas, reducing the risk of security issues. 

\textbf{$\mathbf{RQ_3}$ What are the different characteristics of security-related TD questions?}

\textit{Rationale: }By examining the sentiment of STDQs, the study can gain insight into the attitudes and emotions associated with the STDQs~\cite{sentiment_analysis}. The question score and length of a SO question are indicators of the importance and complexity of the issue being discussed in the question~\cite{asking_about_technical_debt}. Lastly, we identify the tags that require greater community attention to determine whether some questions are more challenging to answer than others~\cite{abdellatif2020challenges}.

\textbf{$\mathbf{RQ_4}$ What are the characteristics of the user profiles asking security-related TD questions?}

\textit{Rationale: }The SO community is known for its diverse users with varying experiences and expertise. Understanding the characteristics of users who ask STDQs, including reputation score, profile age, and badges earned, as previously done by Konstantinos et al.~\cite{stakoulas2022analysis} can help researchers identify patterns and characteristics of users associated with STDQs. This can help inform which users are more likely to address suboptimal security practices in security questions. 

\subsection{Extracting Security Questions}
\label{subsec:extractsq}

In this study, we collected questions and user information from SO by downloading the publicly available SO dump provided by the Stack Exchange Data Dump\footnote{\url{https://archive.org/details/stackexchange}}. Our questions and user dataset spans from 2008 until 2022 and comprises 23,020,127 questions and 19,307,021 users of SO. Each question includes a title and body. 

% Our user dataset spans from 2008 until 2022, containing information for 19,307,021 users of SO.} 
% From that dataset, we used \emph{posts.xml}, containing actual textual content and metadata of the questions. We also use \emph{users.xml} and \emph{badges.xml} to extract the Stack Overflow user information needed to answer our fourth research question.

Since the collected SO questions could be related to any topic, developing a filtering technique to identify the questions related to security was necessary. One strategy would be to use the tags of SO questions and compile all questions with the tag ``security"~\cite{yang2016security}. However, the study by Yang et al.~\cite{yang2016security} revealed that several security questions on SO are not tagged with ``security" but tagged with other security-related terms.  

To overcome this impediment, we followed the guideline of Yang et al.~\cite{yang2016security} to extract the SO questions related to security. The SO questions tagged with ``security," ``sql-injection," ``passwords," ``encryption," ``xss," ``websecurity," ``csrf," ``password-protection," or ``cryptography" will be extracted. Cross-Site Scripting (``xss"), Cross-Site Request Forgery (``csrf"), and ``sql-injection" are related to web vulnerability, while ``cryptography" and ``encryption" are related to cryptography techniques.  

The study of Yang et al.~\cite{yang2016security} independently validated the exclusivity and representativeness of those nine security tags using a two-step heuristic procedure. They started by searching for questions with the tag ``security" in them. In the second step, they extracted the tags from those questions tagged with ``security" and referred to them as candidate tags. Finally, they went through each candidate tag and filtered them out. This selection process resulted in a well-curated set of nine security tags, where the tags were exclusive and representative of the security questions on SO. Using the nine security tags, we extracted 117,233 security-related questions. We used the title and body of the questions for our analysis.

% After identifying 117,233 SO questions related to security, 
Next, we manually examined a sample of 117,233 security questions to assess if the strategy of automated filtering of the questions with the tags led to a high-quality result. In particular, we sampled 383 security questions (with a 95\% confidence level and 5\% margin of error), and two authors independently reviewed and assessed each question to determine whether they were related to security. All disagreements among raters were solved by discussing the questions marked differently by both raters. We used Cohen's Kappa~\cite{cohenkappa} to measure inter-rater reliability. The result showed strong agreement among raters (Cohen's Kappa = 0.89). Our final verification results showed that of the 383 sampled security questions, 370 (97\%) were related to security. We concluded that using the tags by Yang et al.~\cite{yang2016security} is suitable for our study.

\subsection{Classification Model}

After extracting security-related questions, this subsection describes our strategy to identify STDQs. This phase consisted of three main steps, namely (i) preprocessing the extracted security-related questions from the previous section, (ii) training a binary classification model to detect STDQs, and (iii) using the trained classification model to detect STDQs in our preprocessed security-related questions. 

First, we preprocessed our extracted security questions from Section \ref{subsec:extractsq}. This preprocessing step was necessary to clean the extracted security questions. We followed the guidelines by Kozanidis et al.~\cite{asking_about_technical_debt} to preprocess the extracted security questions. We tokenized every question, converted it to lowercase, removed stopwords, and removed all punctuation from the security questions. We also removed any HTML or markdown elements and replaced any instances of source code with a code tag, as these do not include information of lexical or semantic value. We deleted references to ``technical debt," other TD-related keywords suggested by Kozanidis et al.~\cite{asking_about_technical_debt}, and ``SonarQube" from the text to prevent the model from being over-fit during the training phase~\cite{asking_about_technical_debt}.

Next, we trained a binary classification model by following the guidelines by Kozanidis et al.~\cite{asking_about_technical_debt} to detect STDQs. We used their dataset to train our model. Our binary classification model was built using Python's simple transformers library\footnote{\url{https://simpletransformers.ai/}}. We used the pre-trained RoBERTa~\cite{roberta} model as it is considered the state-of-the-art model for binary text classification~\cite{chang2020taming, rajapaksha2021bert}. With the training dataset prepared by Kozanidis et al.~\cite{asking_about_technical_debt}, we used 10-fold cross-validation to train and test the classification model. We used the standard metrics for evaluating our classification model~\cite{rajanbabu2022ensemble, asking_about_technical_debt}, namely precision, recall, and F1 score. With the test set, our model detected STDQs with an F1 score of 0.85.  

Finally, we predicted each prepossessed security question with our trained classifier to identify STDQs. To verify the performance of our classification model, we took a sample with a 95\% confidence level and a 5\% margin of error. Two authors manually verified the output results of the model prediction. By following a labeling guideline of previously published work~\cite{asking_about_technical_debt}, two authors independently reviewed 384 predicted security questions and assessed each question to determine whether it was related to TD. Any disagreements among raters were solved by discussing the questions marked differently by both raters. We had a strong agreement (Cohen's Kappa = 0.81). Our final verification, matched with the model's prediction results, showed that the model achieved an F1 score of 0.75, comparable with the results of the binary classification model of Kozanidis et al.~\cite{asking_about_technical_debt}. 

% We concluded that our classification model is robust enough for classifying TD or non-TD in our security questions.

\begin{table*}[htbp]
\caption{Distribution of STDQs and non-STDQs}
\label{rq2table}
\centering
\begin{tabular}{l lrr lrr}
\toprule
\multicolumn{1}{c}{\multirow{2}{*}{\textbf{Rank}}} & \multicolumn{3}{c}{\textbf{STDQs}} & \multicolumn{3}{c}{\textbf{non-STDQs}} \\ \cmidrule(lr){2-4} \cmidrule(lr){5-7}
\multicolumn{1}{c}{}                      & \textbf{Security Tags}                & \textbf{Questions}  & \textbf{\% per Tag} & \textbf{Security Tags}                & \textbf{Questions}  & \textbf{\% per Tag} \\ \midrule
1                                           & security            & 24,449     & 44.4       & security            & 30,532     & 55.6       \\
2                                           & encryption          & 12,635     & 35.0       & encryption          & 23,396     & 65.0       \\
3                                           & cryptography        & 5,224      & 36.5       & cryptography        & 9,072      & 63.5       \\
4                                           & passwords           & 2,898      & 27.7       & passwords           & 7,555      & 72.3       \\
5                                           & xss                 & 1,910      & 41.8       & csrf                & 2,654      & 58.2       \\
6                                           & csrf                & 1,780      & 41.8       & xss                 & 2,475      & 58.2       \\
7                                           & sql-injection       & 1,582      & 43.0       & sql-injection       & 2,100      & 57.0       \\
8                                           & password-protection & 575        & 33.8       & password-protection & 1,124      & 66.2       \\
9                                           & websecurity         & 150        & 44.0       & websecurity         & 191        & 56.0       \\
\midrule
                                            & Total               & 51,203     & 39.3       & Total               & 79,099     & 60.7          \\ \bottomrule
\end{tabular}
\end{table*}

\subsection{Data Analysis}
\label{subsec:dataanalysis}

We analyzed the questions (both STDQs and non-STDQs) using quantitative and qualitative methods to answer our research questions. For the quantitative analysis, we investigated which of the nine security tags by Yang et al.~\cite{yang2016security} was present in the questions and recorded their frequency. We also examined how many words each question contained to determine whether indications of suboptimal security practices affect a question's length. We considered the post score of a question and the time it took for a question to have an accepted answer. The post score on SO indicates the usefulness of a question as perceived by the community. A higher score means that the post is considered more helpful by the community. The Valence Aware Dictionary and sEntiment Reasoner (VADER) tool were used to analyze the sentiment. VADER is a vocabulary and rule-based sentiment analysis tool and measures the amount of positive or negative emotion and the intensity of emotion in a text~\cite{hutto2014vader}. It is readily available for use on unlabeled text data and is included in Python's NLTK\footnote{\url{www.nltk.org/\_modules/nltk/sentiment/vader.html}} package. We employed VADER to analyze the sentiment of STDQs and non-STDQs since it is a state-of-the-art sentiment analysis tool that performs well in SO setting~\cite{lin2018sentiment}. Both relative sentiment components (negative, neutral, and positive) and compound sentiment scores were used to examine the sentiment expressed in our security questions. According to VADER's GitHub\footnote{\url{www.github.com/cjhutto/vaderSentiment}} site, each word's valence score is added, modified following the guidelines, and normalized to fall between -1 (the most severe negative) and +1 to get the compound score (the most extreme positive). Lastly, we analyzed the user characteristics regarding their reputation score, the years of their account age, the number of profile views, and the badges achieved.

\begin{table*}[htbp]
\caption{Question Types and Definitions}
\label{tab:question_types}
\centering
\begin{tabular}{>{\raggedright\arraybackslash}p{4cm}p{10cm}}
\toprule
\textbf{Question Type}        & \textbf{Definition} \\
\midrule
Debugging                     & Questions for seeking help identifying and resolving errors or bugs in their code. \\
% \addlinespace
Conceptual Understanding      & Questions for comprehending a concept's inner workings or principles. \\
% \addlinespace
Implementation                & Questions for requesting guidance on implementing code features or algorithms. \\
% \addlinespace
Code Review                   & Questions for requesting guidance to assess code quality, readability, and maintainability. \\
% \addlinespace
Best Practices                & Questions for recommendations on adhering to software development best practices. \\
% \addlinespace
Learning Resources            & Questions refer to learning resources to acquire knowledge in a particular subject. \\
\bottomrule
\end{tabular}
\end{table*}

For the qualitative analysis, we took a sample of 396 questions (95\% confidence level and 5\% margin of error) and manually coded the questions. This supplementary coding procedure aims
to understand better the question types and security topics. The \textit{question types} is defined as the nature of the questions and can be categorized into the following: debugging, conceptual understanding, implementation, code review, best practices, and learning resources. Table \ref{tab:question_types} defined each question type. These question types emerged as a result of the provisional coding~\cite{saldana2021coding} from the types reported by Allamanis et al.~\cite{allamanis2013and}. 

% Debugging questions are where users seek help identifying and resolving errors or bugs in their code. Conceptual understanding questions were asked to comprehend a concept's inner workings or principles. Implementation questions requested guidance on implementing specific features or algorithms within their code. Code review questions seek feedback on their code to evaluate its quality, readability, and maintainability. Best practices questions asked for recommendations about adhering to best practices in software development. Learning resource questions refer to various learning resources to acquire knowledge in a particular subject. 

In addition, the \textit{security topics} of the questions are defined as the security topics a question belongs to. This topic differs from the security tags we used to extract security-related questions. Specifically, security tags are provided as SO tags attached to the questions by the users, while security topics emerged from our manual coding process of the questions. The security topics in our security questions were identified using open coding~\cite{saldana2021coding}. One author coded the question types and security topics for each sampled question. Another author reviewed the final classification to ensure the accuracy of the coding procedure. Disagreements were discussed and resolved. \\

\section{Results}
\label{sec:findings}

% This section presents the results organized by RQs. 

\subsection{$\mathbf{RQ_1}$: To what extent do developers indicate suboptimal security practices in security-related questions? }

Out of the 117,233 security-related questions we extracted, our model identified 45,078 (38\%) questions as STDQs. The remaining 72,155 security questions were non-STDQs. However, these 45,078 STDQs do not explicitly mention the word ``technical debt." An example of an STDQ is as follows:

% \begin{quote}
% \emph{``Signed Java Applets have the same security clearance as a normal Java application running on the client. For a particular project, I need these permissions, and I need to perform privileged operations as a result of a JavaScript call. Now, the problem is that, at least for Firefox 3 in Ubuntu (target browser and platform), when an applet method is invoked through unsigned JavaScript it loses its special permissions. As signing the JavaScript is not an option, I need a way to work around this restriction. One way to achieve this is to create a thread when the applet starts, and call methods on that thread whenever the main thread receives the JavaScript calls. I have implemented a working prototype of that idea, but I have found it a bit clumsy, because it uses too much reflection and isn't as easily reusable as I would have wanted."} - Stack Overflow Post ID 1006674\footnote{\url{stackoverflow.com/questions/1006674}}
% \end{quote}

\begin{quote}
    \emph{I use gpg for encrypting a file storing my passwords in Windows. This file is an MS Excel file, which I use for convenience. Every time I want to check or update my passwords ($>$ once per day on average), I execute the following batch script, which decodes the encrypted file and encodes the updated xlsx file again when I close the application. Obviously, this is a suboptimal solution as it creates a decrypted file, which in case of an interruption (e.g. accidentally closing the command line window or a system crash), the file remains unencrypted. Anyone with something better, e.g. using in-memory pipes or the like (in Windows)?} - SO Post ID 66626505\footnote{\url{www.stackoverflow.com/questions/66626505}}
\end{quote}

The question asked for a script for file encryption in Windows. As we can see, although the word ``technical debt" was not mentioned explicitly, this security question clearly discussed suboptimal code solutions, which indicates suboptimal security practices. The person asking the question mentioned how he tried implementing scripts but found them suboptimal.

% \vspace{0.1cm} 
\begin{center}
\fbox{%
\begin{minipage}{0.95\linewidth}
\textbf{Summary $\mathbf{RQ_1}$ (Prevalence of STDQs)} 38\% of security-related questions are STDQs. While addressing suboptimal security issues, developers do not explicitly use the term ``technical debt" but use terminologies indicating suboptimal security practices. 
\end{minipage}}
\end{center}
% \vspace{0.1cm}
% \vspace{-5px}

% \vspace{-5px}
\subsection{$\mathbf{RQ_2}$: Which security tags are more recurrent in security-related TD questions?}

\autoref{rq2table} displays the distribution of security tags in STDQs and non-STDQs. It shows the number of questions per tag and the proportion of STDQs and non-STDQs for each tag. The security tags were ranked in order of most to least frequent number of questions. The ``security" tag appears most frequently in the questions, followed by ``encryption," and in third place, ``cryptography."
% With 24,449 STDQs and 30,532 non-STDQs, the ``security" tag appears most frequently in the questions. The second most popular tag is ``encryption," which includes 12,635 STDQs and 23,396 non-STDQs. 5,224 STDQs and 9,072 non-STDQs were tagged with ``cryptography," which ranks third in frequency. 
Hence, irrespective of STDQs or non-STDQs, the top three tags assigned to the questions are ``security,"  ``encryption," and  ``cryptography."

% The percentages of STDQs in the security tags vary, with 
The security tags with the highest proportion of STDQs are  ``security," ``websecurity," and ``sql-injection" (44.4\%, 44.0\%, and 43.0\%, respectively). At the same time, ``passwords," ``password-protection," and ``encryption" have a lower proportion of STDQs (27.7\%, 33.8\%, and 35.0\%, respectively). Therefore, the ``security," ``websecurity," and ``sql-injection" tags have a higher proportion of STDQs.

Another finding is that the ranking of security tags based on the number of questions for STDQs and non-STDQs is generally identical (we discussed this further in Section \ref{sec:discussion}), except for the tags ``xss" and ``csrf." These two tags have switched positions in the STDQs and non-STDQs, with ``xss" ranking higher in the STDQs and ``csrf" ranking higher in the 
non-STDQs.

% \vspace{0.1cm}
\begin{center}
\fbox{%
\begin{minipage}{0.95\linewidth}
\textbf{Summary $\mathbf{RQ_2}$ (Security Tags Distribution)} The most frequent security tags in STDQs and non-STDQs are ``security," ``encryption," and ``cryptography." The security questions tagged with ``security," ``websecurity," and ``sql-injection" are more likely to indicate suboptimal security practices. The rank of security tags between STDQs and non-STDQs is, in most cases, comparable, with the exception of ``xss" and ``csrf" tags.
\end{minipage}}
\end{center}
% \vspace{0.1cm}

\subsection{$\mathbf{RQ_3}$: What are the different characteristics of security-related TD questions?}

The characteristics of the STDQs and non-STDQs are compared in \autoref{rq3table}. The table lists the features (e.g., question length, compound score, and post score), mean and median scores across STDQs and non-STDQs.

\begin{table}[htbp]
\caption{Comparative Analysis of Different Aspects Between TD and non-TD Security-Related Questions}
\label{rq3table}
\centering
\begin{tabular}{l rr rr}
\toprule
\multirow{2}{*}{} & \multicolumn{2}{c}{\textbf{STDQs}}   & \multicolumn{2}{c}{\textbf{non-STDQs}}  \\ 
\cmidrule(lr){2-3} \cmidrule(lr){4-5}
                  & \textbf{Mean}   & \textbf{Median} & \textbf{Mean}   & \textbf{Median} \\ 
\midrule
Question Length (Words)  & 132.39 & 112.00 & 79.29  & 67.00  \\
Negative Score    & 0.04   & 0.03   & 0.04   & 0.03   \\
Positive Score    & 0.10   & 0.09   & 0.09   & 0.08   \\
Neutral Score     & 0.86   & 0.86   & 0.86   & 0.87   \\
Compound Score    & 0.43   & 0.69   & 0.26   & 0.40   \\
Post Score        & 3.47   & 1.00   & 2.44   & 1.00   \\
Time to Answer (Hours) & 378.22 & 1.19   & 391.03 & 1.07   \\
\bottomrule
\end{tabular}
\end{table}

Our study's results show that the median question length of STDQs is 112 words. Conversely, the non-STDQs are characterized by shorter median question lengths with 67 words. This indicates that STDQs are lengthier than non-STDQs.

The median neutral sentiment scores were 0.86 and 0.87 for the STDQs and non-STDQs, respectively. Positive sentiment in both questions has lower scores (median sentiment score = 0.09), while negative sentiment is even lower (median sentiment score = 0.03). The median compound score (the normalized sum of positive, neutral, and negative scores) for STDQs is 0.69, higher than the median for non-STDQs (0.40). Overall, the sentiment in STDQs and non-STDQs are comparable, indicating almost no difference in sentiment from users when asking security questions regardless of the presence of suboptimal security practices indications.

The mean post score (to indicate the usefulness of a SO question post) for STDQs is 3.47, and the median score is 1.00. In contrast, non-STDQs have a mean post score of 2.44 and a median score 1.00. This shows that the median post score of STDQs and non-STDQs is the same, indicating the same attention from SO users.

Finally, the median time to answer STDQs is 1.19 hours, while the median time needed to answer non-STDQs is 1.07 hours. According to these findings, the time for a question to have an accepted answer is comparable for the two questions. 

% This result indicated the same challenges in answering TD and non-TD security questions.

% \vspace{0.1cm}
\begin{center}
\fbox{%
\begin{minipage}{0.95\linewidth}
\textbf{Summary $\mathbf{RQ_3}$ (Different Aspects of Questions)} Compared to non-STDQs, STDQs require more words. The median compound score for STDQs is higher compared to non-STDQs. STDQs have a slightly higher mean post score and higher median time to answer.
\end{minipage}}
\end{center}
% \vspace{0.1cm}

\subsection{$\mathbf{RQ_4}$: What are the user profiles asking security-related TD questions?}

\begin{table}[htbp]
\caption{Comparative Analysis of User Profiles Asking TD and non-TD Security-Related Questions}
\label{rq4table}
\centering
\begin{tabular}{l rr rr}
\toprule
\multirow{2}{*}{} & \multicolumn{2}{c}{\textbf{STDQs}}   & \multicolumn{2}{c}{\textbf{non-STDQs}}  \\ 
\cmidrule(lr){2-3} \cmidrule(lr){4-5}
                  & \textbf{Mean}   & \textbf{Median} & \textbf{Mean}   & \textbf{Median} \\ 
\midrule
User Profile Age          & 9      & 10     & 8      & 9      \\
Reputation Score  & 4377   & 469    & 2597   & 193    \\
Profile Views     & 478    & 59     & 276    & 36     \\
Badges            & Silver & Bronze & Silver & Bronze \\
\bottomrule
\end{tabular}
\end{table}

The results for ${RQ_4}$ are presented in \autoref{rq4table}. It compares the user profiles of those who asked STDQs and non-STDQs, showing the mean and median values for account age, reputation score, profile views, and each group's most common badge type.

The median account age for asking STDQs is ten years old. In contrast, those who asked non-STDQs had a slightly lower median account age of 9 years. Users who asked STDQs had a higher median reputation score, with 469 scores for STDQs and 193 scores for non-STDQs. Similar patterns emerged in profile views, with STDQs users reporting greater median (59 views) profile views than non-STDQs users (36 views). Bronze is the most prevalent (median) badge type for both categories.

% The results for ${RQ_4}$ suggest that users asking STDQs have more experience, higher reputation scores, and more profile views than users asking non-STDQs. However, the badge distribution is identical between both groups.

% \vspace{0.1cm}
\begin{center}
\fbox{%
\begin{minipage}{0.95\linewidth}
\textbf{Summary $\mathbf{RQ_4}$ (User Profiles)} Compared to non-STDQs, users who ask STDQs tend to be more experienced, have higher reputation ratings, and have more profile views. Both groups have a similar distribution of badges.
\end{minipage}}
\end{center}
% \vspace{0.1cm}

\subsection{Qualitative Analysis: Question Types}

\begin{figure*}[!h]
\centering
\includegraphics[width=18cm]{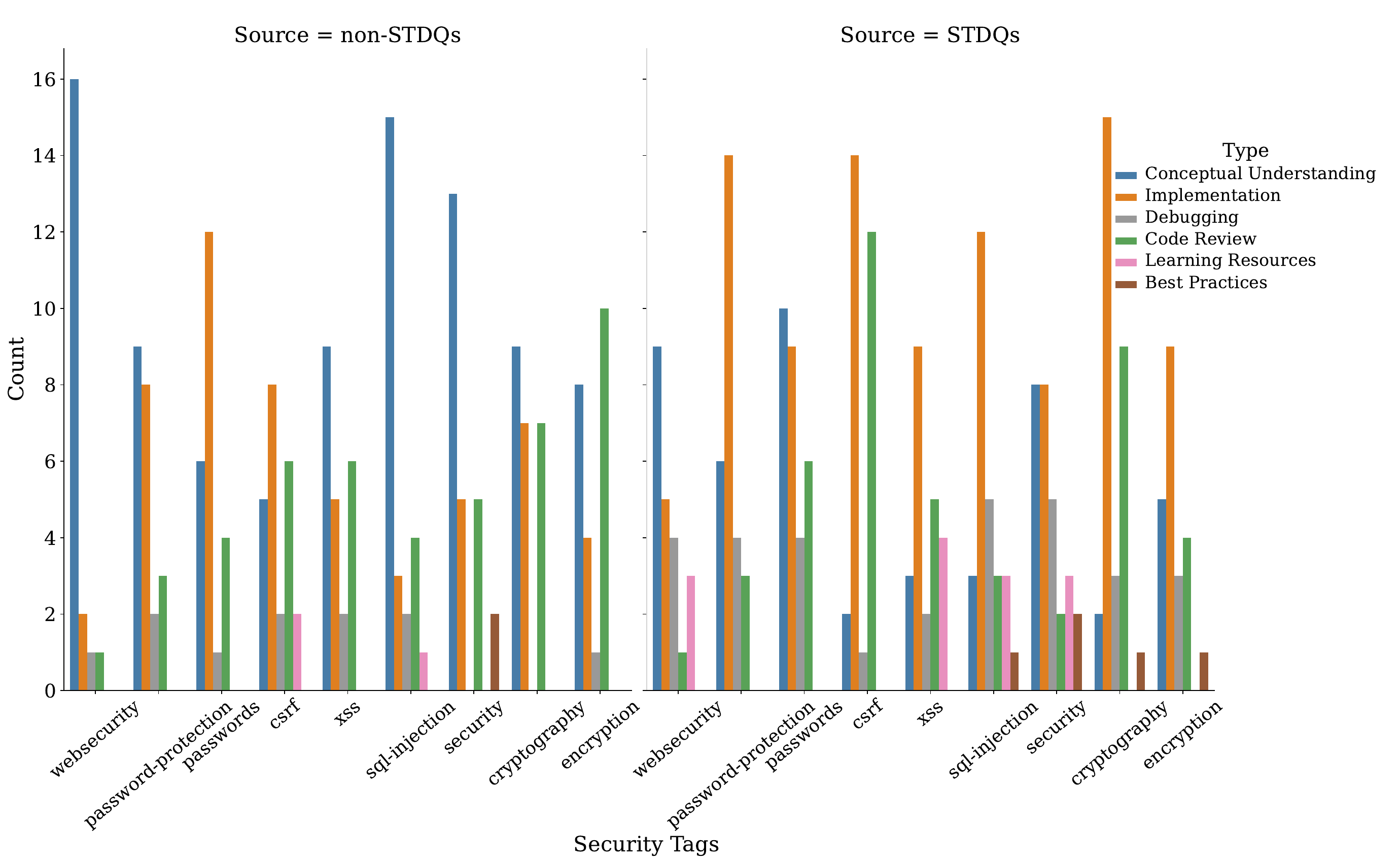}
\caption[Caption for fig1]{Distribution of Question Types for each Security Tag}
\label{fig:TDQuestionTypes}
\end{figure*}

% \begin{figure}[!h]
% \centering
% \includegraphics[width=8.7cm]{Figures/Question Types NTD.pdf}
% \caption[Caption for fig1]{Question Types for non-\textcolor{red}{STDQs}}
% \label{fig:NTDQuestionTypes}
% \end{figure}

\vspace{-1.5px}
\begin{table*}[htbp]
\caption{Comparison of Question Types between TD and non-TD Security-Related Questions}
\label{questiontypeslabel}
\centering
\begin{tabular}{l l r l r}
\toprule
\multirow{2}{*}{\textbf{Rank}} & \multicolumn{2}{c}{\textbf{STDQs}}   & \multicolumn{2}{c}{\textbf{non-STDQs}}  \\
\cmidrule(lr){2-3} \cmidrule(lr){4-5}
                      & \textbf{Type}                     & \textbf{No. of Questions} & \textbf{Type}                     & \textbf{No. of Questions} \\ 
\midrule
1                     & Implementation           &  95                   & Conceptual Understanding & 90                   \\
2                     & Conceptual Understanding & 48                   & Implementation           & 54                   \\
3                     & Debugging                & 45                   & Debugging                & 46                   \\
4                     & Best Practices           & 31                   & Best Practices           & 11                   \\
5                     & Code Review              & 13                   & Code Review              & 3                    \\
6                     & Learning Resources       & 5                    & Learning Resources       & 2                    \\
Total                 &                           & 237                  &                           & 206                  \\
\bottomrule
\end{tabular}
\end{table*}

In this section, we report the results of the question types, which we described in Section \ref{subsec:dataanalysis} as the nature of STDQs and non-STDQs obtained through provisional coding~\cite{saldana2021coding} from the types given by Allamanis et al.~\cite{allamanis2013and}. The distribution of question types (defined in Table \ref{tab:question_types}) for STDQs and non-STDQs is shown in \autoref{questiontypeslabel}. Since a security question can have multiple question types, we can assign multiple question-type labels to the question. However, the ranking of the question types for STDQs and non-STDQs differs. Most STDQs are of ``Implementation" type. Contrarily, ``Conceptual Understanding" is the most common type for non-STDQs. Other question types like ``Debugging," ``Best Practices," ``Code Review," and ``Learning Resources" are less common in both questions.

Figure \ref{fig:TDQuestionTypes} further summarizes our analysis of question types per security tag. Most security tags in STDQs have ``Implementation" as their question types, except for ``websecurity" and ``passwords." In contrast, most security tags in non-STDQs have ``Conceptual Understanding" as the question type, except for ``passwords," ``csrf," and ``encryption."

% \vspace{0.1cm}
\begin{center}
\fbox{%
\begin{minipage}{0.95\linewidth}
\textbf{Summary Question Types} Implementation is the most discussed question type in STDQs. In contrast, conceptual knowledge questions are the main focus of non-STDQs. Other question types are less frequent in both STDQs and non-STDQs.
\end{minipage}}
\end{center}
% \vspace{0.1cm}

\subsection{Qualitative Analysis: Security Topics}

\begin{table}[htbp]
\caption{Comparison of Security Topics between TD and non-TD Security-Related Questions}
\label{securitytopics}
\centering
\begin{tabular}{l l rr}
\toprule
\multirow{2}{*}{\textbf{Security Tags}} & \multirow{2}{*}{\textbf{Security Topics}} & \multicolumn{2}{c}{\textbf{No. of Questions}}  \\
\cmidrule(lr){3-4}
                               &                                 & \textbf{STDQs}      & \textbf{non-STDQs}      \\ 
\midrule
\multicolumn{1}{l}{csrf}      & Prevention                      & 19      & 8        \\
                               & Token                           & 9       & 13       \\
                               & Concept                         & 2       & 2        \\ 
\midrule
\multicolumn{1}{l}{sql-injection} & Detection                   & 2       & 5        \\
                                  & Concept                     & 4       & 8        \\
                                  & Prevention                  & 22      & 7        \\
                                  & Simulation                  & 0       & 3        \\ 
\midrule
\multicolumn{1}{l}{xss}        & Detection                       & 1       & 1        \\
                                & Prevention                      & 17      & 7        \\
                                & Testing                         & 0       & 3        \\
                                & Concept                         & 2       & 5        \\
                                & Simulation                      & 0       & 3        \\ 
\midrule
\multicolumn{1}{l}{cryptography} & Signature                    & 6       & 11       \\
                                  & Algorithm                    & 22      & 7        \\
                                  & Key                          & 1       & 0        \\ 
\midrule
\multicolumn{1}{l}{encryption}  & Algorithm                      & 9       & 18       \\
                                 & File                           & 14      & 8        \\
                                 & Server and Client Side         & 6       & 6        \\
                                 & Key                            & 4       & 3        \\
                                 & IP Address                     & 0       & 2        \\
                                 & Kubernetes                     & 0       & 4        \\
                                 & Mobile                         & 1       & 2        \\
\bottomrule
\end{tabular}
\end{table}

In this section, we report the results of the security topics described in Section \ref{subsec:dataanalysis} as the topics of STDQs and non-STDQs obtained through the open coding process. These security topics are different from the SO security tags. A comparison of the distribution of security topics for STDQs and non-STDQs within each security tag is summarized in \autoref{securitytopics}. Due to page limitations, we only show the results for some of the security topics.

The distribution of security topics for STDQs and non-STDQs in each security tag differs, indicating that the two types of inquiries have different priorities. The emphasis on practical implementations is typically more significant in STDQs, e.g., ``Prevention" techniques for ``csrf," ``sql-injection," and ``xss." Non-STDQs focus more on conceptual elements, e.g., ``Token" for ``csrf," and have a more uniform distribution across topics.

Non-STDQs with the ``cryptography" tag emphasize ``Signature" more than STDQs, which prioritize ``Algorithm." Similarly, STDQs with the ``encryption" tag focus more on ``File." In contrast, non-STDQs are more focused on ``Algorithm."

\vspace{-2px}
\begin{center}
\fbox{%
\begin{minipage}{0.95\linewidth}
\textbf{Summary Security Topics} Each security tag emphasized STDQs and non-STDQs differently. ``Prevention" in ``csrf," ``xss," and ``sql-injection" is highlighted in STDQs. ``Algorithm" and ``File" are given priority in STDQs for ``cryptography" and ``encryption."
\end{minipage}}
\end{center} 

\vspace{4px}

\section{Discussion}
\label{sec:discussion}

\textbf{Prevalence of TD in Security-Related Questions.}
Previous studies investigated TD-related questions on SO~\cite{asking_about_technical_debt, alfayez2023asked}. In our study, we specifically observed STDQs. Developers are voicing concerns about potentially poor implementations or design choices as they solve various security-related problems. Our results also reveal that developers only implicitly convey their concerns about suboptimal security practices during the software development process rather than mentioning ``technical debt" explicitly when asking security-related questions.

Indications of suboptimal security practices in security questions can be attributed to the nature of the SO website and the intertwined relationship between security and TD. Developers use SO to improve their code quality and ensure best practices are being followed~\cite{an2017stack}. At the same time, previous studies~\cite{siavvas2019empirical, nord2016can, izurieta2018position, oishwee2022exploratory, sultana2020examining} showed that the relationship between suboptimal implementations or design choices in software systems is strongly correlated with security vulnerabilities. The nature of the SO website and the intertwined relationship result in developers asking STDQs to gain insights and find effective solutions, albeit suboptimal security practices, which are implicitly (and potentially unconsciously) indicated in the security questions.

The absence of explicit TD mentions in all STDQs can be attributed to a lack of awareness among developers of the term TD. The latter is still unfamiliar to almost 45\% developers, as suggested by Rama{\v{c}} et al.~\cite{ramavc2020common}. As per the studies by Kozanidis et al.~\cite{asking_about_technical_debt} and Alfayez et al.~\cite{alfayez2023asked}, the low number (only 500) of questions on SO questions, irrespective of domain, explicitly mentioning TD, confirms that the TD metaphor is unfamiliar to many practitioners. 

% In addition, the study by Kozanidis et al.~\cite{asking_about_technical_debt} and Alfayez et al.~\cite{alfayez2023asked} supports this statement, as only about 500 questions on SO questions, irrespective of domain, mention TD explicitly. 

\textbf{Security Topics in STDQs and non-STDQs.}
\autoref{rq2table} displays the distribution of security tags in STDQs and non-STDQs. The ranking of security tags for STDQs and non-STDQs is almost identical. One potential reason can be attributed to the overlapping concerns between TD and security in software. Since many security-related decisions involve trade-offs between functionality, maintainability, and security, security-related issues involve overlapping concerns with TD~\cite{siavvas2019empirical, nord2016can, izurieta2018position, oishwee2022exploratory, sultana2020examining}. This overlap contributes to the similarity in the results between STDQs and non-STDQs, as developers need to address the same topics when managing security, regardless of the presence of TD.

% Another potential reason could be associated with the consistency in developer challenges. The similar findings between TD and non-TD questions might imply that developers face challenges across both categories in similar security areas. Whether or not developers are specifically discussing TD, this consistency suggests that security-related concerns are fundamentally complicated or challenging to implement~\cite{lopez2019anatomy}.}

% According to our quantitative investigation, 
The ``encryption" and ``cryptography" security-related tags are the second and third most frequently appearing in STDQs and non-STDQs. This appears to be caused by the fact that ``cryptography" and ``encryption" are among the most popular security tags on SO~\cite{lopez2018investigation, yang2016security}, regardless of whether the queries are TD or non-TD related. The complexity of cryptography principles, which have been known to present difficulties for developers attempting to understand them~\cite{hazhirpasand2021hurdles}, could be another reason for their predominance. Our qualitative analysis of the security topics for ``cryptography" and ``encryption" tags revealed a prominent lack of understanding of developing cryptography algorithms and understanding concepts related to signatures and file encryption. Our findings align with the findings of the study of Hazhirpasand et al.~\cite{hazhirpasand2021hurdles}, which also reported those security topics as the main hurdles in Cryptography for developers.

A more in-depth qualitative examination of the data revealed that web vulnerabilities associated with three security tags, namely ``xss," ``csrf," and ``sql-injection" have significantly more STDQs. Additionally, our analysis shows that STDQs with those tags emphasize addressing preventative measures. Previous studies~\cite{shalini2011prevention, kombade2012client, kar2013prevention} show that implementing effective prevention strategies is challenging for developers. This difficulty makes them discuss suboptimal solutions and design choices, resulting in more STDQs.\\

% non-TD questions focusing on XSS, CSRF, and SQL Injection typically cover a broader spectrum of subjects, such as vulnerability detections, simulations, testing, and other concepts. In contrast, 

\textbf{Characteristics of Questions and Users.}
By quantitatively comparing various aspects of STDQs and non-STDQs, we observed that STDQs tend to use more words than non-STDQs. From a subsequent qualitative analysis, we note that STDQs mainly request guidance on implementing specific features, algorithms, or techniques within the code written by the users, thus requiring more words and details to describe the context. In contrast, non-STDQs mainly ask short, general, and open-ended questions to comprehend the inner workings or principles behind a concept, language, or technology. As a result, non-STDQs are shorter, as they do not need as many words to elaborate on the details or need to include code. We conclude that STDQs are more complex and challenging. Non-STDQs instead focus primarily on conceptual understanding, thus requiring fewer words. We observe that the median word count in our STDQs is comparable with the results of the study by Kozanidis et al.~\cite{asking_about_technical_debt} and the standard length of SO questions~\cite{calefato2018ask}. Our analysis of STDQs length reveals a consistent pattern across numerous research efforts, highlighting the underlying complexity of STDQs. 

% By being prepared to handle these complex issues, developers can make more informed decisions when tackling security challenges and managing technical debt.}

Our quantitative analysis also shows that users asking STDQs have higher reputation scores and more profile views. This result can be attributed to the nature of STDQs, as they are more complex and challenging to comprehend for newer users. Thus, higher reputation users are more aware of incurring TD and the potential implications of such actions on security issues and are more likely to include suboptimal security issues in their questions.\\

% \vspace{2px}
\section{Implications}
\label{sec:implications}
\textbf{For researchers.} The findings of our study can be used to pinpoint the security areas that are highly intertwined with TD. Knowledge of these areas allows researchers to prioritize research and analysis, especially in critical areas of the security domain. Our results reveal that web vulnerability tags (``xss," ``csrf," and ``sql-injection") are prominent since they are associated with a higher proportion of STDQs than most security tags. This is primarily due to a need for knowledge of prevention techniques and indicates the need for a more in-depth investigation into developers' need for knowledge of web vulnerability prevention techniques. Leaving the developers to struggle with these web vulnerability prevention techniques will cause severe consequences to clients and companies~\cite{shalini2011prevention, kombade2012client, kar2013prevention}. Researchers and tool developers should focus on creating more comprehensive, user-friendly tools and methodologies for implementing web vulnerability prevention techniques. These tools can guide implementing prevention techniques to minimize or eliminate TD without compromising software security.

\textbf{For practitioners.} Practitioners with more experience are more aware and likely to participate in discussions about suboptimal security practices. The lack of awareness from less experienced developers urges more experienced developers to help less experienced team members understand how their decisions in security may affect incurring suboptimal security practices. In order to increase the quality of the software-generated, experienced developers should mentor and direct less experienced developers in recognizing and managing suboptimal security practices by highlighting their potential implications on security issues. 

\textbf{For educators.} Our study found a lack of explicit mention of the term ``technical debt" in STDQs. This result means there is a lack of awareness among developers of the term TD in the security domain. Thus, security courses and seminars should be structured to educate about suboptimal security practices and TD concepts. With better-structured courses and seminars, we will raise the awareness of developers of TD and any suboptimal practices in the security domain to reduce security vulnerabilities in software development. \\

% Moreover, we found numerous security-related questions on Stack Overflow related to TD, particularly in areas like cryptography, encryption, and web vulnerabilities (XSS, CSRF, and SQL Injection). Developers in these areas may need more significant resources and training to understand the effective implementation of security and prevent incurring TD in the software. By setting priorities, developers can reduce TD and subsequent security vulnerabilities resulting from less-than-ideal implementations in these fields.}

\section{Threats to Validity}
\label{sec:threats_to_validity}

Despite our best efforts, validity threats could impact the findings of this study. Following Runeson et al.~\cite{threats_to_validity} classification, we consider the following factors to discuss the potential risks to this study and the relevant mitigation measures taken.

\textbf{Construct Validity} is regarding how well our operational measures are suited to respond to our RQs. Our datasets of security-related questions were extracted based on a set of predefined security tags by Yang et al.~\cite{yang2016security}. Since the SO tagging system is user-generated, there may be inaccuracies and inconsistencies in how posts are tagged in the used dataset. To mitigate this threat, two authors manually inspected a statistically significant sample of the security question posts, as described in Section~\ref{subsec:extractsq}, to verify the content. The manual validation stages could be a potential threat to our study. Although we manually validated the classification model predictions, the process might have been prone to human error and bias. To mitigate this threat, two authors labeled the sampled datasets by following the guidelines prepared by Kozanidis et al.~\cite{asking_about_technical_debt}. Disagreements among raters were resolved through discussion until an agreement was reached. 

\textbf{Internal Validity} is regarding how much the ``treatment," and not other factors, are responsible for the observed findings. The most relevant threat to the internal validity of our study regards the classification model used to identify STDQs. The model trained to identify STDQs is based on the RoBERTa model~\cite{roberta}, which achieved an F1 score of 0.75. This suggests the dataset may contain false positives and negatives that could influence the outcomes. Future research can reduce this risk by enhancing model performance with more training datasets. 

\textbf{External Validity} is regarding the scope of generalizability of our findings. In this study, the analyzed data was collected using SO. Therefore, the findings from our study might lack generalizability, as there are other platforms, such as Software Engineering Stack Exchange\footnote{\url{www.softwareengineering.stackexchange.com}}, where developers share their thoughts about security and TD. However, as also stated in the study by Kozanidis et al.~\cite{asking_about_technical_debt}, among the well-known group of Stack Exchange Question and Answer (Q\&A) sites, SO is the most used programming Q\&A website with over 24M questions and 20M registered users. Given its size and popularity, we deem SO as the best pool of data where developers discuss topics related to technical debt and security vulnerabilities in software development. Another threat to our study lies in using the dataset presented by Kozanidis et al.~\cite{asking_about_technical_debt} to train our classification model. In their study, the authors documented that they deliberately constructed their automated query as encompassing as possible to filter out only the TD-related question posts on SO. Nevertheless, their findings might not reflect all TD-related queries featured on SO. Since we use their TD-related question post dataset to train our model, this threat is also cascaded to our study. We can reduce this threat in future studies by developing our own dataset.

\textbf{Conclusion Validity} is regarding the link between outcomes and our treatments. Our dataset of security-related questions was extracted based on a set of predefined security tags by Yang et al.~\cite{yang2016security}. However, there might be more security-related tags than those nine security tags. This threat will impact the number of extracted security-related questions, thus also impacting the results and conclusions of our study. Additionally, the model trained to identify STDQs only achieved an F1 score of 0.75. The model was imperfect, suggesting the identification of STDQs to contain false positives and negatives that could influence the results and conclusions of our study. The manual coding analyses could be another threat to our study. We inspected a statistically significant random sample from the STDQs and non-STDQs to better comprehend the security topics and question types in the dataset. Sampling bias might be present, causing the sampled STDQs and non-STDQs not to represent the whole population. Additionally,  the manual coding procedure might incorporate the subjective judgment of the authors. Despite the collaborative review and editing process between the two authors, there could still be biases or inconsistencies that affect the conclusions.

% The manual validation stages could be regarded as a potential threat to the conclusion validity of our study. Although we manually validated the classification model predictions, the process might have been prone to human error and bias. To mitigate this threat, two authors labeled the sampled datasets by following the guidelines prepared by Kozanidis et al.~\cite{asking_about_technical_debt}. Disagreements among raters were resolved through discussion between the two authors until an agreement was reached. Additionally, while the qualitative analysis aims to better comprehend the security topics and question types in the dataset, the manual coding procedure might incorporate the subjective judgment of the authors. Despite the collaborative review and editing process between the two authors, there could still be biases or inconsistencies that affect the conclusions.
\vspace{-2px}
\section{Conclusions and Future Work}
\label{conclusion}

In this study, we explored the different characteristics of STDQs. We used a deep-learning approach to identify STDQs. The classification model revealed that 38\% of security questions on SO are STDQs. The most recurrent tags among the STDQs emerged as ``security" and ``encryption." The latter typically have a neutral sentiment, are lengthier, and are posed by users with higher reputation scores. 

For future works, we plan to extend the scope of our analysis to other platforms, such as Github or other Q\&A platforms. Analyzing data from other platforms would provide a more comprehensive understanding of the relationship between TD and security. Another interesting avenue of research is to enhance our model's capability in detecting STDQs by expanding the training datasets and using more sophisticated models. Lastly, we would like to perform further correlation studies to study the relationship between STDQs and user profiles.

% , we investigated the extent to which developers discuss TD in the security domain, identify recurrent security tags in STDQs, examine various aspects of STDQs, and analyze the user profiles associated with STDQs. 

\section*{Acknowledgment}
This research is partly supported by an NSERC Collaborative Research and Training Experience (CREATE) grant on Software Analytics at the University of Saskatchewan and the European Union under the Italian National Recovery and Resilience Plan (NRRP) of NextGenerationEU, partnership on Telecommunications of the Future” (PE0000001 - program “RESTART”).

% The preferred spelling of the word ``acknowledgment'' in America is without 
% an ``e'' after the ``g''. Avoid the stilted expression ``one of us (R. B. 
% G.) thanks $\ldots$''. Instead, try ``R. B. G. thanks$\ldots$''. Put sponsor 
% acknowledgments in the unnumbered footnote on the first page.

\bibliographystyle{IEEEtran}
\bibliography{references}

% Generated by IEEEtran.bst, version: 1.14 (2015/08/26)
\begin{thebibliography}{10}
\providecommand{\url}[1]{#1}
\csname url@samestyle\endcsname
\providecommand{\newblock}{\relax}
\providecommand{\bibinfo}[2]{#2}
\providecommand{\BIBentrySTDinterwordspacing}{\spaceskip=0pt\relax}
\providecommand{\BIBentryALTinterwordstretchfactor}{4}
\providecommand{\BIBentryALTinterwordspacing}{\spaceskip=\fontdimen2\font plus
\BIBentryALTinterwordstretchfactor\fontdimen3\font minus
  \fontdimen4\font\relax}
\providecommand{\BIBforeignlanguage}[2]{{%
\expandafter\ifx\csname l@#1\endcsname\relax
\typeout{** WARNING: IEEEtran.bst: No hyphenation pattern has been}%
\typeout{** loaded for the language `#1'. Using the pattern for}%
\typeout{** the default language instead.}%
\else
\language=\csname l@#1\endcsname
\fi
#2}}
\providecommand{\BIBdecl}{\relax}
\BIBdecl

\bibitem{wen2019learning}
\BIBentryALTinterwordspacing
S.-F. Wen and B.~Katt, ``Learning software security in context: An evaluation
  in open source software development environment,'' in \emph{Proceedings of
  the 14th International Conference on Availability, Reliability and Security},
  ser. ARES '19.\hskip 1em plus 0.5em minus 0.4em\relax New York, NY, USA:
  Association for Computing Machinery, 2019. [Online]. Available:
  \url{https://doi.org/10.1145/3339252.3340336}
\BIBentrySTDinterwordspacing

\bibitem{assal2018security}
H.~Assal and S.~Chiasson, ``Security in the software development lifecycle.''
  in \emph{SOUPS@ USENIX Security Symposium}, 2018, pp. 281--296.

\bibitem{yang2016security}
X.-L. Yang, D.~Lo, X.~Xia, Z.-Y. Wan, and J.-L. Sun, ``What security questions
  do developers ask? a large-scale study of stack overflow posts,''
  \emph{Journal of Computer Science and Technology}, vol.~31, pp. 910--924,
  2016.

\bibitem{moradian2013security}
E.~Moradian, ``Security of e-commerce software systems,'' \emph{Agent and
  Multi-Agent Systems in Distributed Systems-Digital Economy and E-Commerce},
  pp. 95--103, 2013.

\bibitem{khan2022systematic}
R.~A. Khan, S.~U. Khan, H.~U. Khan, and M.~Ilyas, ``Systematic literature
  review on security risks and its practices in secure software development,''
  \emph{ieee Access}, vol.~10, pp. 5456--5481, 2022.

\bibitem{khan2011secure}
R.~Khan, ``Secure software development: a prescriptive framework,''
  \emph{Computer Fraud \& Security}, vol. 2011, no.~8, pp. 12--20, 2011.

\bibitem{michael2006security}
H.~Michael and L.~Steve, ``The security development lifecycle: Sdl: A process
  for developing demonstrably more secure software,'' 2006.

\bibitem{geer2010companies}
D.~Geer, ``Are companies actually using secure development life cycles?''
  \emph{Computer}, vol.~43, no.~6, pp. 12--16, 2010.

\bibitem{tondel2008security}
I.~A. Tondel, M.~G. Jaatun, and P.~H. Meland, ``Security requirements for the
  rest of us: A survey,'' \emph{IEEE software}, vol.~25, no.~1, pp. 20--27,
  2008.

\bibitem{kruchten2012technical}
P.~Kruchten, R.~L. Nord, and I.~Ozkaya, ``Technical debt: From metaphor to
  theory and practice,'' \emph{IEEE Software}, vol.~29, no.~6, pp. 18--21,
  2012.

\bibitem{alves2016identification}
\BIBentryALTinterwordspacing
N.~S. Alves, T.~S. Mendes, M.~G. de~Mendon\c{c}a, R.~O. Sp\'{\i}nola, F.~Shull,
  and C.~Seaman, ``Identification and management of technical debt,''
  \emph{Inf. Softw. Technol.}, vol.~70, no.~C, p. 100–121, feb 2016.
  [Online]. Available: \url{https://doi.org/10.1016/j.infsof.2015.10.008}
\BIBentrySTDinterwordspacing

\bibitem{lim2012balancing}
E.~Lim, N.~Taksande, and C.~Seaman, ``A balancing act: What software
  practitioners have to say about technical debt,'' \emph{IEEE Software},
  vol.~29, no.~6, pp. 22--27, 2012.

\bibitem{tom2013exploration}
\BIBentryALTinterwordspacing
E.~Tom, A.~Aurum, and R.~Vidgen, ``An exploration of technical debt,'' \emph{J.
  Syst. Softw.}, vol.~86, no.~6, p. 1498–1516, jun 2013. [Online]. Available:
  \url{https://doi.org/10.1016/j.jss.2012.12.052}
\BIBentrySTDinterwordspacing

\bibitem{mcgraw2012software}
G.~McGraw, ``Software security: Building security in,'' \emph{Datenschutz und
  Datensicherheit-DuD}, vol.~36, no.~9, pp. 662--665, 2012.

\bibitem{siavvas2018static}
M.~Siavvas, E.~Gelenbe, D.~Kehagias, and D.~Tzovaras, ``Static analysis-based
  approaches for secure software development,'' in \emph{Security in Computer
  and Information Sciences: First International ISCIS Security Workshop 2018,
  Euro-CYBERSEC 2018, London, UK, February 26-27, 2018, Revised Selected Papers
  1}.\hskip 1em plus 0.5em minus 0.4em\relax Springer International Publishing,
  2018, pp. 142--157.

\bibitem{siavvas2019optimum}
M.~Siavvas and E.~Gelenbe, ``Optimum checkpoints for programs with loops,''
  \emph{Simulation Modelling Practice and Theory}, vol.~97, p. 101951, 2019.

\bibitem{siavvas2019empirical}
M.~Siavvas, D.~Tsoukalas, M.~Jankovic, D.~Kehagias, A.~Chatzigeorgiou,
  D.~Tzovaras, N.~Anicic, and E.~Gelenbe, ``An empirical evaluation of the
  relationship between technical debt and software security,'' in \emph{9th
  International Conference on Information society and technology (ICIST)}, vol.
  2019, 2019.

\bibitem{kuhn2018can}
R.~Kuhn, M.~Raunak, and R.~Kacker, ``Can reducing faults prevent
  vulnerabilities?'' \emph{Computer}, vol.~51, no.~7, pp. 82--85, 2018.

\bibitem{camilo2015bugs}
F.~Camilo, A.~Meneely, and M.~Nagappan, ``Do bugs foreshadow vulnerabilities? a
  study of the chromium project,'' in \emph{Conf. on MSR}, 2015, pp. 269--279.

\bibitem{izurieta2018position}
C.~Izurieta, D.~Rice, K.~Kimball, and T.~Valentien, ``A position study to
  investigate technical debt associated with security weaknesses,'' in
  \emph{Proceedings of the 2018 International Conference on technical debt},
  2018, pp. 138--142.

\bibitem{siavvas2022technical}
M.~Siavvas, D.~Tsoukalas, M.~Jankovic, D.~Kehagias, and D.~Tzovaras,
  ``Technical debt as an indicator of software security risk: a machine
  learning approach for software development enterprises,'' \emph{Enterprise
  Information Systems}, vol.~16, no.~5, p. 1824017, 2022.

\bibitem{rindell2019security}
K.~Rindell and J.~Holvitie, ``Security risk assessment and management as
  technical debt,'' in \emph{2019 International Conference on Cyber Security
  and Protection of Digital Services (Cyber Security)}.\hskip 1em plus 0.5em
  minus 0.4em\relax IEEE, 2019, pp. 1--8.

\bibitem{alfayez2023asked}
R.~Alfayez, Y.~Ding, R.~Winn, G.~Alfayez, C.~Harman, and B.~Boehm, ``What is
  asked about technical debt (td) on stack exchange question-and-answer (q\&a)
  websites? an observational study,'' \emph{Empirical Software Engineering},
  vol.~28, no.~2, p.~35, 2023.

\bibitem{asking_about_technical_debt}
\BIBentryALTinterwordspacing
N.~Kozanidis, R.~Verdecchia, and E.~Guzman, ``Asking about technical debt:
  Characteristics and automatic identification of technical debt questions on
  stack overflow,'' in \emph{Proceedings of the 16th ACM / IEEE International
  Symposium on Empirical Software Engineering and Measurement}, ser. ESEM
  '22.\hskip 1em plus 0.5em minus 0.4em\relax New York, NY, USA: Association
  for Computing Machinery, 2022, p. 45–56. [Online]. Available:
  \url{https://doi.org/10.1145/3544902.3546245}
\BIBentrySTDinterwordspacing

\bibitem{barua2014developers}
A.~Barua, S.~W. Thomas, and A.~E. Hassan, ``What are developers talking about?
  an analysis of topics and trends in stack overflow,'' \emph{Empirical
  Software Engineering}, vol.~19, pp. 619--654, 2014.

\bibitem{silva2021topic}
C.~C. Silva, M.~Galster, and F.~Gilson, ``Topic modeling in software
  engineering research,'' \emph{Empirical Software Engineering}, vol.~26,
  no.~6, p. 120, 2021.

\bibitem{rosen2016mobile}
C.~Rosen and E.~Shihab, ``What are mobile developers asking about? a large
  scale study using stack overflow,'' \emph{Empirical Software Engineering},
  vol.~21, pp. 1192--1223, 2016.

\bibitem{haque2020challenges}
M.~U. Haque, L.~H. Iwaya, and M.~A. Babar, ``Challenges in docker development:
  A large-scale study using stack overflow,'' in \emph{Proceedings of the 14th
  ACM/IEEE International Symposium on Empirical Software Engineering and
  Measurement (ESEM)}, 2020, pp. 1--11.

\bibitem{gama2020using}
E.~Gama, S.~Freire, M.~Mendon{\c{c}}a, R.~O. Sp{\'\i}nola, M.~Paixao, and M.~I.
  Cort{\'e}s, ``Using stack overflow to assess technical debt identification on
  software projects,'' in \emph{Proceedings of the XXXIV Brazilian Symposium on
  Software Engineering}, 2020, pp. 730--739.

\bibitem{croft2022empirical}
R.~Croft, Y.~Xie, M.~Zahedi, M.~A. Babar, and C.~Treude, ``An empirical study
  of developers’ discussions about security challenges of different
  programming languages,'' \emph{Empirical Software Engineering}, vol.~27, pp.
  1--52, 2022.

\bibitem{lopez2019anatomy}
T.~Lopez, T.~Tun, A.~Bandara, L.~Mark, B.~Nuseibeh, and H.~Sharp, ``An anatomy
  of security conversations in stack overflow,'' in \emph{2019 IEEE/ACM 41st
  International Conference on Software Engineering: Software Engineering in
  Society (ICSE-SEIS)}.\hskip 1em plus 0.5em minus 0.4em\relax IEEE, 2019, pp.
  31--40.

\bibitem{gqm}
V.~R. B.~G. Caldiera and H.~D. Rombach, ``The goal question metric approach,''
  \emph{Encyclopedia of software engineering}, pp. 528--532, 1994.

\bibitem{sentiment_analysis}
W.~Medhat, A.~Hassan, and H.~Korashy, ``Sentiment analysis algorithms and
  applications: A survey,'' \emph{Ain Shams engineering journal}, vol.~5,
  no.~4, pp. 1093--1113, 2014.

\bibitem{abdellatif2020challenges}
\BIBentryALTinterwordspacing
A.~Abdellatif, D.~Costa, K.~Badran, R.~Abdalkareem, and E.~Shihab, ``Challenges
  in chatbot development: A study of stack overflow posts,'' in
  \emph{Proceedings of the 17th International Conference on Mining Software
  Repositories}, ser. MSR '20.\hskip 1em plus 0.5em minus 0.4em\relax New York,
  NY, USA: Association for Computing Machinery, 2020, p. 174–185. [Online].
  Available: \url{https://doi.org/10.1145/3379597.3387472}
\BIBentrySTDinterwordspacing

\bibitem{stakoulas2022analysis}
\BIBentryALTinterwordspacing
K.~Stakoulas, K.~Georgiou, N.~Mittas, and L.~Angelis, ``An analysis of user
  profiles from covid-19 questions in stack overflow,'' in \emph{25th
  Pan-Hellenic Conference on Informatics}, ser. PCI 2021.\hskip 1em plus 0.5em
  minus 0.4em\relax New York, NY, USA: Association for Computing Machinery,
  2022, p. 419–424. [Online]. Available:
  \url{https://doi.org/10.1145/3503823.3503900}
\BIBentrySTDinterwordspacing

\bibitem{cohenkappa}
J.~Cohen, ``A coefficient of agreement for nominal scales,'' \emph{Educational
  and psychological measurement}, vol.~20, no.~1, pp. 37--46, 1960.

\bibitem{roberta}
Y.~Liu, M.~Ott, N.~Goyal, J.~Du, M.~Joshi, D.~Chen, O.~Levy, M.~Lewis,
  L.~Zettlemoyer, and V.~Stoyanov, ``Roberta: A robustly optimized bert
  pretraining approach,'' \emph{arXiv preprint arXiv:1907.11692}, 2019.

\bibitem{chang2020taming}
\BIBentryALTinterwordspacing
W.-C. Chang, H.-F. Yu, K.~Zhong, Y.~Yang, and I.~S. Dhillon, ``Taming
  pretrained transformers for extreme multi-label text classification,'' in
  \emph{Proceedings of the 26th ACM SIGKDD International Conference on
  Knowledge Discovery \&amp; Data Mining}, ser. KDD '20.\hskip 1em plus 0.5em
  minus 0.4em\relax New York, NY, USA: Association for Computing Machinery,
  2020, p. 3163–3171. [Online]. Available:
  \url{https://doi.org/10.1145/3394486.3403368}
\BIBentrySTDinterwordspacing

\bibitem{rajapaksha2021bert}
P.~Rajapaksha, R.~Farahbakhsh, and N.~Crespi, ``Bert, xlnet or roberta: the
  best transfer learning model to detect clickbaits,'' \emph{IEEE Access},
  vol.~9, pp. 154\,704--154\,716, 2021.

\bibitem{rajanbabu2022ensemble}
K.~Rajanbabu, I.~K. Veetil, V.~Sowmya, E.~Gopalakrishnan, and K.~Soman,
  ``Ensemble of deep transfer learning models for parkinson's disease
  classification,'' in \emph{Soft Computing and Signal Processing: Proceedings
  of 3rd ICSCSP 2020, Volume 2}.\hskip 1em plus 0.5em minus 0.4em\relax
  Springer, 2022, pp. 135--143.

\bibitem{hutto2014vader}
C.~Hutto and E.~Gilbert, ``Vader: A parsimonious rule-based model for sentiment
  analysis of social media text,'' in \emph{Proceedings of the international
  AAAI conference on web and social media}, vol.~8, no.~1, 2014, pp. 216--225.

\bibitem{lin2018sentiment}
B.~Lin, F.~Zampetti, G.~Bavota, M.~Di~Penta, M.~Lanza, and R.~Oliveto,
  ``Sentiment analysis for software engineering: How far can we go?'' in
  \emph{Proceedings of the 40th international conference on software
  engineering}, 2018, pp. 94--104.

\bibitem{saldana2021coding}
J.~Salda{\~n}a, ``The coding manual for qualitative researchers,'' \emph{The
  coding manual for qualitative researchers}, pp. 1--440, 2021.

\bibitem{allamanis2013and}
M.~Allamanis and C.~Sutton, ``Why, when, and what: analyzing stack overflow
  questions by topic, type, and code,'' in \emph{2013 10th Working conference
  on mining software repositories (MSR)}.\hskip 1em plus 0.5em minus
  0.4em\relax IEEE, 2013, pp. 53--56.

\bibitem{an2017stack}
L.~An, O.~Mlouki, F.~Khomh, and G.~Antoniol, ``Stack overflow: A code
  laundering platform?'' in \emph{2017 IEEE 24th International Conference on
  Software Analysis, Evolution and Reengineering (SANER)}.\hskip 1em plus 0.5em
  minus 0.4em\relax IEEE, 2017, pp. 283--293.

\bibitem{nord2016can}
R.~L. Nord, I.~Ozkaya, E.~J. Schwartz, F.~Shull, and R.~Kazman, ``Can knowledge
  of technical debt help identify software vulnerabilities?'' in \emph{CSET@
  USENIX Security Symposium}, 2016.

\bibitem{oishwee2022exploratory}
S.~J. Oishwee, Z.~Codabux, and N.~Stakhanova, ``An exploratory study on the
  relationship of smells and design issues with software vulnerabilities,'' in
  \emph{Proceedings of the 1st International Workshop on Mining Software
  Repositories Applications for Privacy and Security}, 2022, pp. 16--20.

\bibitem{sultana2020examining}
K.~Z. Sultana, Z.~Codabux, and B.~Williams, ``Examining the relationship of
  code and architectural smells with software vulnerabilities,'' in \emph{2020
  27th Asia-Pacific Software Engineering Conference (APSEC)}.\hskip 1em plus
  0.5em minus 0.4em\relax IEEE, 2020, pp. 31--40.

\bibitem{ramavc2020common}
R.~Rama{\v{c}}, V.~Mandi{\'c}, N.~Tau{\v{s}}an, N.~Rios, M.~G.
  de~Mendonca~Neto, C.~Seaman, and R.~O. Sp{\'\i}nola, ``Common causes and
  effects of technical debt in serbian it: Insightd survey replication,'' in
  \emph{2020 46th euromicro conference on software engineering and advanced
  applications (seaa)}.\hskip 1em plus 0.5em minus 0.4em\relax IEEE, 2020, pp.
  354--361.

\bibitem{lopez2018investigation}
T.~Lopez, T.~T. Tun, A.~Bandara, M.~Levine, B.~Nuseibeh, and H.~Sharp, ``An
  investigation of security conversations in stack overflow: Perceptions of
  security and community involvement,'' in \emph{Proceedings of the 1st
  international workshop on security awareness from design to deployment},
  2018, pp. 26--32.

\bibitem{hazhirpasand2021hurdles}
M.~Hazhirpasand, O.~Nierstrasz, M.~Shabani, and M.~Ghafari, ``Hurdles for
  developers in cryptography,'' in \emph{2021 IEEE International Conference on
  Software Maintenance and Evolution (ICSME)}.\hskip 1em plus 0.5em minus
  0.4em\relax IEEE, 2021, pp. 659--663.

\bibitem{shalini2011prevention}
S.~Shalini and S.~Usha, ``Prevention of cross-site scripting attacks (xss) on
  web applications in the client side,'' \emph{International Journal of
  Computer Science Issues (IJCSI)}, vol.~8, no.~4, p. 650, 2011.

\bibitem{kombade2012client}
R.~Kombade and B.~Meshram, ``Client side csrf defensive tool,''
  \emph{International Journal of Information and Network Security}, vol.~1,
  no.~3, p. 171, 2012.

\bibitem{kar2013prevention}
D.~Kar and S.~Panigrahi, ``Prevention of sql injection attack using query
  transformation and hashing,'' in \emph{2013 3rd IEEE International Advance
  Computing Conference (IACC)}.\hskip 1em plus 0.5em minus 0.4em\relax IEEE,
  2013, pp. 1317--1323.

\bibitem{calefato2018ask}
F.~Calefato, F.~Lanubile, and N.~Novielli, ``How to ask for technical help?
  evidence-based guidelines for writing questions on stack overflow,''
  \emph{Information and Software Technology}, vol.~94, pp. 186--207, 2018.

\bibitem{threats_to_validity}
P.~Runeson and M.~H{\"o}st, ``Guidelines for conducting and reporting case
  study research in software engineering,'' \emph{Empirical software
  engineering}, vol.~14, pp. 131--164, 2009.

\end{thebibliography}
\end{document}